\documentclass[sigconf]{acmart}
\AtBeginDocument{%
  \providecommand\BibTeX{{%
    \normalfont B\kern-0.5em{\scshape i\kern-0.25em b}\kern-0.8em\TeX}}}

\usepackage{algorithm}
\usepackage{algorithmic}
\usepackage{bm}
\usepackage{colortbl}
\usepackage{arydshln}
\usepackage{multirow}
\usepackage{multicol}
\usepackage{xcolor}

\usepackage{enumitem}

\copyrightyear{2021}
\acmYear{2021}
\setcopyright{acmcopyright}\acmConference[ICAIF'21]{2nd ACM International Conference on AI in Finance}{November 3--5, 2021}{Virtual Event, USA}
\acmBooktitle{2nd ACM International Conference on AI in Finance (ICAIF'21), November 3--5, 2021, Virtual Event, USA}
\acmPrice{15.00}
\acmDOI{10.1145/3490354.3494366}
\acmISBN{978-1-4503-9148-1/21/11}

\acmConference[New York '21]{New York '20: ACM International Conference on AI in Finance}{Nov. 3--5, 2021}{New York, NY}
\begin{document}

\title{FinRL: Deep Reinforcement Learning Framework to Automate Trading in Quantitative Finance}

\author{Xiao-Yang Liu, Hongyang Yang}
%\authornote{Co-primary author with equal contribution.}
\email{{xl2427, hy2500}@columbia.edu}
\affiliation{%
  \institution{Columbia University}
  %\city{New York City}
  %\state{New York}
}

%\author{Hongyang Yang}
%\email{hy2500@columbia.edu}
%\affiliation{%
%  \institution{AI4Finance Foundation}
  %\city{New York City}
  %\state{New York}
%}

\author{Jiechao Gao}
\email{jg5ycn@virginia.edu}
\affiliation{%
  \institution{University of Virginia}
  %\city{Bellevue}
  %\state{Washington}
}

\author{Christina Dan Wang}
\authornote{Corresponding author.}
\email{christina.wang@nyu.edu}
\affiliation{%
  \institution{New York University Shanghai}
  %\city{Murray Hill}
  %\state{New Jersey}
}

\begin{abstract}
  Deep reinforcement learning (DRL) has been envisioned to have a competitive edge in quantitative finance. However, there is a steep development curve for quantitative traders to obtain an agent that automatically positions to win in the market, namely \textit{to decide where to trade, at what price} and \textit{what quantity}, due to the error-prone programming and arduous debugging. In this paper, we present the first open-source framework \textit{FinRL} as a full pipeline to help quantitative traders overcome the steep learning curve. FinRL is featured with simplicity, applicability and extensibility under the key principles, \textit{full-stack framework, customization, reproducibility} and \textit{hands-on tutoring}. 
  
  Embodied as a three-layer architecture with modular structures, FinRL implements fine-tuned state-of-the-art DRL algorithms and common reward functions, while alleviating the debugging workloads. Thus, we help users pipeline the strategy design at a high turnover rate. At multiple levels of time granularity, FinRL simulates various markets as training environments using historical data and live trading APIs. Being highly extensible, FinRL reserves a set of user-import interfaces and incorporates trading constraints such as market friction, market liquidity and investor's risk-aversion. Moreover, serving as practitioners' stepping stones, typical trading tasks are provided as step-by-step tutorials, e.g., stock trading, portfolio allocation, cryptocurrency trading, etc.
  
  % cryptocurrency trading, 
  
  %Furthermore, we include many Jupyter notebooks as step-by-step tutorials, namely stock trading, cryptocurrency trading, and portfolio allocation. The FinRL library is available on Github. %\url{https://github.com/AI4Finance-LLC/FinRL-Library}.
\end{abstract}

%\ccsdesc[500]{Computing methodologies~Policy iteration}
%\ccsdesc[500]{Computing methodologies~Value iteration}

\maketitle

% \section1{Introduction}
%\usepackage{soul}
\vspace{-0.2in}
\section{Introduction}

% Many problems in stock market are decision problems with sequential nature, like option pricing, portfolio optimization, risk management, and deep reinforcement learning is a natural solution method.
% Deep reinforcement learning is a natural solution to some stock trading problems. However, it is challenging to apply deep reinforcement learning model in practice since it calls for complex algorithm constructions and accessible training environments. In this paper, we propose a library function containing deep reinforcement learning applications in stock market. The library function provides environments based on DJIA, S&P 500 and Nasdaq 100. It also provided classic deep reinforcement learning algorithms and demos which can be called directly from the command line. The proposed library can easily help users to realize deep reinforcement learning algorithms and make comparisons.

% \cite{ Nan2020SentimentAK, zhang2020deep, Ganesh2018DeepRL}

\begin{figure}[t]
\centering
\includegraphics[height=5cm]{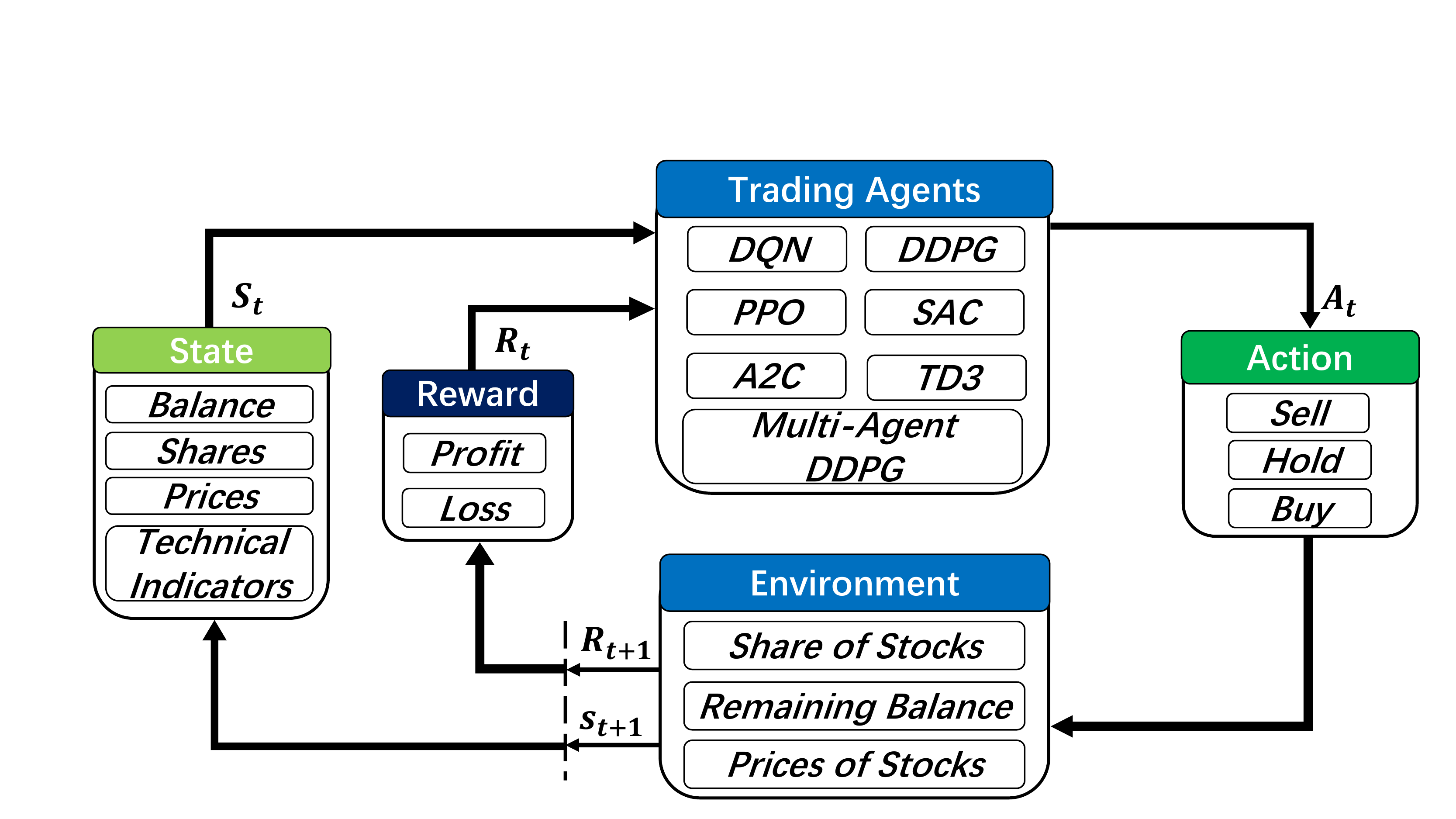}\vspace{-0.2in}
\caption{Overview of automated trading in FinRL, using deep reinforcement learning.}\vspace{-0.2in}
\label{state_transition}
\vspace{-0.1in}
\end{figure}

Deep reinforcement learning (DRL), that balances exploration (of uncharted territory) and exploitation (of current knowledge), is a promising approach to automate trading in quantitative finance \cite{xiong2018practical}\cite{yang2020}\cite{vadori2020risk}\cite{zhang2020deep}\cite{Jiang2017ADR}\cite{RL_survey}. DRL algorithms are powerful in solving dynamic decision making problems by learning through interactions with an unknown environment, and offer two major advantages of \textit{portfolio scalability} and \textit{market model independence} \cite{buehler2019deep}. In quantitative finance, algorithmic trading is essentially making dynamic decisions, namely \textit{to decide where to trade, at what price and what quantity}, in a highly stochastic and complex financial market. Incorporating many financial factors, as shown in Fig. \ref{state_transition}, a DRL trading agent builds a multi-factor model to trade automatically, which is difficult for human traders to accomplish \cite{bekiros2010fuzzy,zhang2017online}. Therefore, DRL has been envisioned to have a competitive edge in quantitative finance.

%\yanglet{To Chen, There are many more works using DRL. Check the ICAIF 2020 website: \url{https://ai-finance.org/conference-program/} and also previous finance workshop at NeurIPS 2018, 2019, ICML 2019, KDD 2019, etc. and Google them.}

Many existing works have applied DRL in quantitative financial tasks. Both researchers and industry practitioners are actively designing trading strategies fueled by DRL, since deep neural networks are significantly powerful at estimating the expected return of taking a certain action at a state. Moody and Saffell \cite{moody2001learning} utilized a policy search for stock trading; Deng \textit{et al.} \cite{deng2016deep} showed that DRL can obtain more profits than conventional methods. 
More applications include stock trading \cite{Nan2020SentimentAK, zhang2020deep,vadori2020risk,yang2020},
futures contracts \cite{zhang2020deep},
alternative data (news sentiments) \cite{Nan2020SentimentAK,Koratamaddi2021MarketSD},
high frequency trading \cite{Ganesh2018DeepRL}, 
liquidation strategy analysis \cite{bao2019multi},
and hedging \cite{buehler2019deep}.
DRL is also being actively explored in the cryptocurrency market, e.g., automated trading, portfolio allocation, and market making.

%\yanglet{challenges}.

%\yanglet{1. Why the "training-testing" workflow in machine learning fall short for financial tasks?}

However, designing a DRL trading strategy is not easy. The programming is error-prone with tedious debugging. The development pipeline includes preprocessing market data, building a training environment, managing trading states, and backtesting trading performance. These steps are standard for implementation but yet time consuming especially for beginners. Therefore, there is an urgent demand for an open-source library to help researchers and quantitative traders to overcome the steep learning curve. 

 %such as transactions cost, market liquidity and the investor's degree of risk-aversion. , where users can better test the robustness of their strategy with  The environment is represented by action space and state space

%\yanglet{To Chen, Besides all the good functionalities, let us also itemize the three features: completeness, hands-on tutorials and reproducibility, to attract beginners.}

%处理金融task更便捷，

%{\color{red}why we need FinRL? 

%tensorflow,pytorch weakness in financial area, 

%tensorflow and pytorch not user-friendly to finance area

%突出个性化-finrl在金融领域中相比于TF和Pytorch的独特之处 }

%In this paper, we present a three-layer \textit{FinRL} framework that automatically streamlines the development of trading strategies and aims to help researchers and quantitative traders to iterate their strategies at a high turnover rate. FinRL provides common building blocks \bruce{what are common building blocks} that allow strategy builders to configure datasets or APIs as virtual environments, to train deep neural networks as trading agents, to analyze trading performance via extensive backtesting, and to incorporate market frictions.
In this paper, we present a \textit{FinRL} framework that automatically streamlines the development of trading strategies, so as to help researchers and quantitative traders to iterate their strategies at a high turnover rate.
%目的是帮助researchers和trader去快速迭代策略
%这部分的逻辑应该是：FinRL怎么帮助快速迭代策略，researchers和trader只需提供configrations然后就能看结果了
Users specify the configurations, such as picking data APIs and DRL algorithms, and analyze the performance of trading results. %the automated backtesting module.
%三层架构是什么
To achieve this, FinRL introduces a \textit{three-layer} framework. At the bottom is an environment layer that simulates financial markets using actual historical data, such as closing price, shares, trading volume, and technical indicators. In the middle is the agent layer that implements fine-tuned DRL algorithms and common reward functions. The agent interacts with the environment through properly defined reward functions on the state space and action space. The top layer includes applications in automated trading, where we demonstrate several use cases, namely stock trading, portfolio allocation, cryptocurrency trading, etc. We provide baseline trading strategies to alleviate debugging workloads. 
%The top layer includes multifarious applications in automated trading, such as stock trading and portfolio allocation. 

%Those applications are plug-and-play for users.

% (DQN \cite{mnih2015human}, DDPG \cite{lillicrap2015continuous}, multi-agent DDPG \cite{lowe2017multi}, PPO \cite{schulman2017proximal}, SAC \cite{haarnoja2018soft}, A2C \cite{mnih2016asynchronous} and TD3 \cite{Dankwa2019TwinDelayedDA}, etc.)

%The DRL agents include DQN \cite{mnih2015human}, DDPG \cite{lillicrap2015continuous}, Adaptive DDPG \cite{li2019optimistic}, Multi-Agent DDPG \cite{lowe2017multi}, PPO \cite{schulman2017proximal}, SAC \cite{haarnoja2018soft}, A2C \cite{mnih2016asynchronous} and TD3 \cite{fujimoto2018addressing} algorithms.
%做框架，服务于自动化流程，帮助用户节省时间和精力，加快迭代过程
%1. 降低学习曲线
%2. 提供自动化的数据处理，回测，可供选择的算法，目标函数等
%3. 帮助策略开发，让用户可以快速迭代过程，尽快收敛出有机会使用的交易策略。

Under the three-layer framework, FinRL is developed with three primary principles:
\begin{itemize}[leftmargin=*]
    %三点principles没有解释清楚
    %\item \textbf{Completeness}. We provide a complete DRL framework thatis customized and optimized for financial trading tasks. Users can transparently make use of the development pipeline.
    \item \textbf{Full-stack framework}. To provide a full-stack DRL framework with finance-oriented optimizations, including market data APIs, data preprocessing, DRL algorithms, and automated backtesting. Users can transparently make use of such a development pipeline.
    %\item \textbf{Reproducibility}. Benchmark schemes for typical trading tasks provide users stepping stones. 
    %包现有各种经典agents实现且容易引入新agent算法，可以组合这些算法去产生新策略，通过简单配置可以reproduce的。因此容易按需复现各种策略。
    \item \textbf{Customization}. To maintain modularity and extensibility in development by including state-of-the-art DRL algorithms and supporting design of new algorithms. The DRL algorithms can be used to construct trading strategies by simple configurations. %Thus, users can use FinRL to reproduce strategies on demand.
    % \item \textbf{Hands-on tutoring}. We provide beginner-friendly tutorials to help users walk through the pipeline of a strategy design while exploring the functionalities and features.
    \item \textbf{Reproducibility and hands-on tutoring}. To provide tutorials such as step-by-step Jupyter notebooks and user guide to help users walk through the pipeline and reproduce the use cases. 
\end{itemize}
This leads to a unified framework where developers are able to efficiently explore ideas through high-level configurations and specifications, and to customize their own strategies at request.

Our contributions are summarized as follows:
\begin{itemize}[leftmargin=*]
  \item FinRL is the first open-source framework that demonstrates the great potential of applying DRL algorithms in quantitative finance. We build an ecosystem around the FinRL framework, which seeds the rapidly growing AI4Finance community.
  %Meanwhile, FinRL supports training on a cloud, since the modules are packaged inside containers.
  %Typical trading tasks are implemented as benchmarks, and hands-on tutorials are provided in a beginner-friendly and reproducible fashion. % using Jupyter notebooks. 
  %It implements several benchmark trading tasks that can be reproduced by beginners.
    \item The application layer provides interfaces for users to customize FinRL to their own trading tasks. Automated backtesting module and performance metrics are provided to help quantitative traders iterate trading strategies at a high turnover rate. Profitable trading strategies are reproducible and hands-on tutorials are provided in a beginner-friendly fashion. Adjusting the trained models to the rapid changing markets is also possible.
  \item The agent layer provides state-of-the-art DRL algorithms that are adapted to finance with fine-tuned hyperparameters. Users can add new DRL algorithms. %Therefore, via the agent layer, the trading strategies are reproducible.
  %\item Adjusting the trained models to the rapid changing FinRL supports users to rapidly adjust the trained models to the changing markets. Because both training environment and DRL agent can be set via simple configurations. Besides, training environment and DRL agent configurations
    \item The environment layer includes not only a collection of historical data APIs, but also live trading APIs. They are reconfigured into standard OpenAI gym-style environments \cite{brockman2016openai}. Moreover, it incorporates market frictions and allows users to customize the trading time granularity. 

  %has good scalability since state-of-the-art DRL algorithms can be trained on a cloud computing platform. Automatic backtesting module and performance metrics are provided to help quantitative traders develop trading strategies at a high turnover rate. Adjusting the trained models to the rapid changing markets is well-supported.   FinRL provides hands-on tutorials in a beginner-friendly fashion.
\end{itemize}

The remainder of this paper is organized as follows. Section 2 reviews related works. Section 3 presents the FinRL framework. Section 4 demonstrates benchmark trading tasks using FinRL. We conclude this paper in Section 5.

% \section2{ExistingWorks} 
%Policy-based methods offer a few advantages over value-prediction methods like DQN presented in the previous three Posts. One is that, as we already discussed, we no longer have to worry about devising an action-selection strategy like ϵ-greedy policy; instead, we directly sample actions from the policy. And this is important; remember that we wasted a lot of time fixing up methods to improve the stability of training our DQN. For instance, we had to use experience replay and target networks, and there are several other methods in the academic literature that helps. A policy network tends to simplify some of that complexity.

\section{Related Works}

We review the state-of-the-art DRL algorithms, relevant open-source libraries, and applications of DRL in quantitative finance.  

\subsection{Deep Reinforcement Learning Algorithms}

%https://towardsdatascience.com/a-beginners-guide-to-q-learning-c3e2a30a653c
%https://towardsdatascience.com/crystal-clear-reinforcement-learning-7e6c1541365e
%https://towardsdatascience.com/policy-based-methods-8ae60927a78d
%https://towardsdatascience.com/deep-reinforcement-learning-for-automated-stock-trading-f1dad0126a02
%https://stackoverflow.com/questions/37370015/what-is-the-difference-between-value-iteration-and-policy-iteration

Many DRL algorithms have been developed. They fall into three categories: value based, policy based, and actor-critic based.

A value based algorithm estimates a state-action value function that guides the optimal policy. Q-learning \cite{watkins1992q} approximates a Q-value (expected return) by iteratively updating a Q-table, which works for problems with small discrete state spaces and action spaces. Researchers proposed to utilize deep neural networks for approximating Q-value functions, e.g., deep Q-network (DQN) and its variants double DQN and dueling DQN \cite{SpinningUp2018}.

%e.g., deep Q-network (DQN) \cite{lillicrap2015continuous, mnih2015human} and its variants double DQN \cite{Hasselt2016DeepRL} and dueling DQN \cite{Wang2016DuelingNA}.

%DQN algorithms are good at handling discrete state spaces, such trading with a single stock.

A policy based algorithm directly updates the parameters of a policy through policy gradient \cite{Sutton00policygradient}. Instead of value estimation, policy gradient uses a neural network to model the policy directly, whose input is a state and output is a probability distribution according to which the agent takes an action at the input state. %The advantage of policy gradient is that it can handle continuous action spaces, such as continuous price and portfolio allocation.

An actor-critic based algorithm combines the advantages of value based and policy based algorithms. It updates two neural networks, namely, an actor network updates the policy (probability distribution) while a critic network estimates the state-action value function. During the training process, the actor network takes actions and the critic network evaluates those actions. The state-of-art actor-critic based algorithms are deep deterministic policy gradient (DDPG), proximal policy optimization (PPO), asynchronous advantage actor critic (A3C), advantage actor critic (A2C), soft actor-critic (SAC), multi-agent DDPG, and twin-delayed DDPG (TD3) \cite{SpinningUp2018}.

%The state-of-art actor-critic based algorithms are deep deterministic policy gradient (DDPG) \cite{lillicrap2015continuous}, proximal policy optimization (PPO) \cite{schulman2017proximal}, asynchronous advantage actor critic (A3C) \cite{a3c_2016}, advantage actor critic (A2C) \cite{a3c_2016}, soft actor-critic (SAC) \cite{haarnoja2018soft}, multi-agent DDPG \cite{lowe2017multi}, and twin-delayed DDPG (TD3) \cite{Dankwa2019TwinDelayedDA}.

%FinRL has consolidated and elaborated upon those algorithms to build financial DRL \& RL models.  {\color{red}However, }

\begin{figure*}[t]
\centering
\includegraphics[height=8cm]{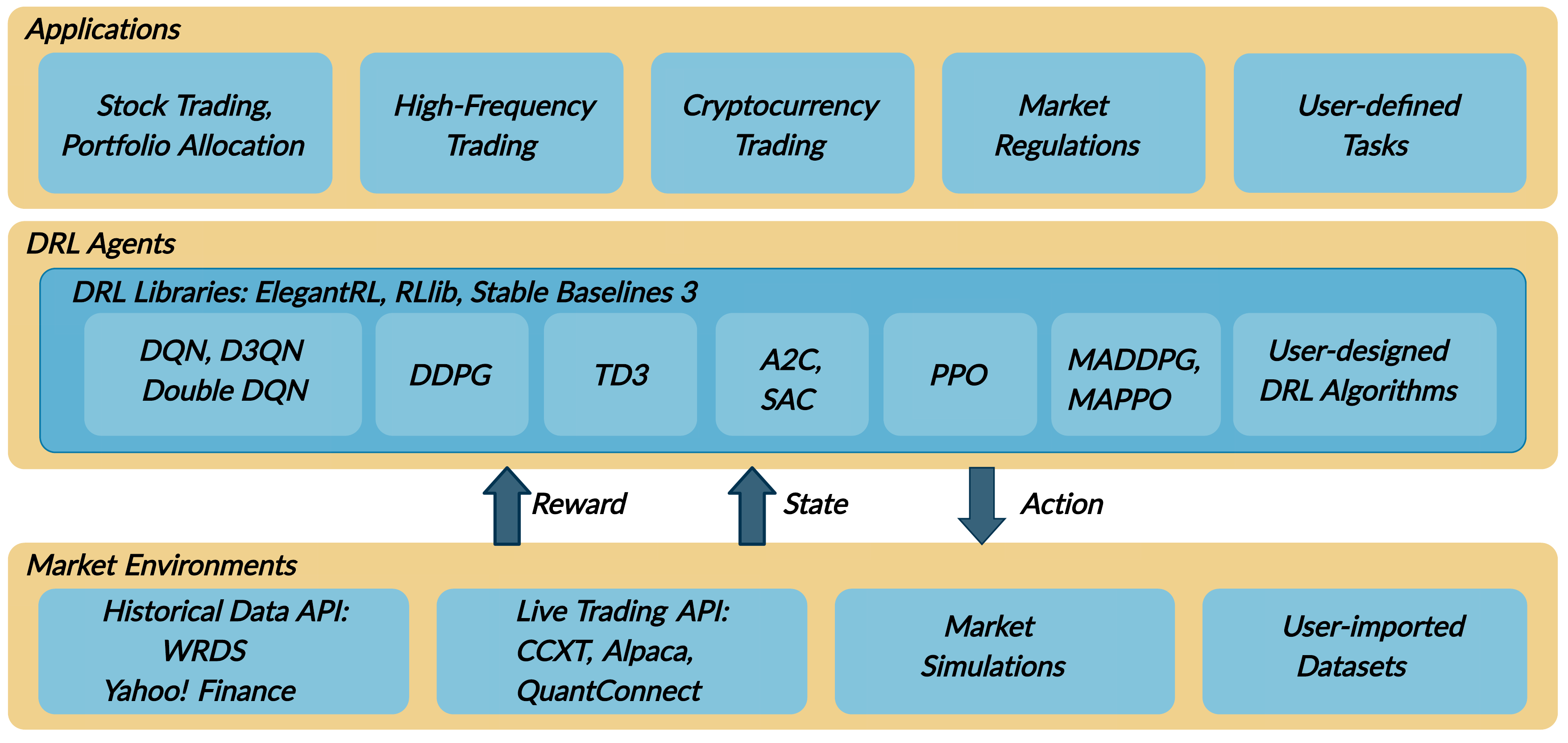}\vspace{-0.2in}
\caption{Overview of FinRL: application layer at the top, agent layer in the middle and environment layer at the bottom.}
\label{overview_finrl}
\vspace{-0.1in}
\end{figure*}

%historical data APIs and live trading APIs are wrapped up as an environment layer at the bottom.

\subsection{Deep Reinforcement Learning Libraries}
%\yanglet{Do we keep the library section or not? If merging with Section 2.1, it will be very long.}
%\begin{table*}[htb]
%\caption{Comparison of RL Libraries \yanglet{what about: environments %(finance), statistics, tutorials, reproducibility (use cases), etc.}}
%\begin{tabular}{|l|l|l|l|l|}
%\hline
%\textbf{Libraries}       & \textbf{SOTA RL}        & \textbf{Doc.}           & \textbf{Pip Installer}        & \textbf{Finance Customization}       \\ \hline
%\textbf{FinRL}       & Yes     & Yes     & Yes     & Yes         \\ \hline
%\textbf{OpenAI Gym}       & Yes     & No     & Yes     & No         \\ \hline
%\textbf{Google Dopamine}         & Yes     & Yes     & Yes     & No         \\ \hline
%\textbf{RLlib}          & Yes     & Yes     & Yes     & No         \\ \hline
%\textbf{RLlab}             & Yes     & Yes     & Yes     & No         \\ \hline
%\textbf{Personae}         & Yes     & No     & Yes     & No         \\ \hline
%\textbf{Tensorforce}        & Yes     & Yes     & Yes     & No         \\ \hline
%\textbf{Horizon}        & Yes     & No     & No     & No         \\ \hline
%\end{tabular}
%\end{table*}
%In Table 1 we compare them from different perspectives with our library.

%There are a number of open-source DRL libraries. 

We summarize relevant open-source DRL libraries as follows:

\noindent \textbf{OpenAI Gym} \cite{brockman2016openai} provides standardized environments for various DRL tasks. OpenAI baselines \cite{baselines} implements common DRL algorithms, while stable baselines 3 \cite{stable-baselines} improves \cite{baselines} with code cleanup and user-friendly examples.

\noindent \textbf{Google Dopamine} \cite{castro18dopamine} aims for fast prototyping of DRL algorithms. It features good plugability and reusability. 

\noindent \textbf{RLlib} \cite{liang2018rllib} provides highly scalable DRL algorithms. It has modular framework and is well maintained.

%\item \textbf{PyBrain} \cite{schaul2010pybrain} is a modular machine learning library. It contains algorithms for supervised/unsupervised learning, RL and blackbox optimizations. 

\noindent \textbf{TensorLayer} \cite{dong2017tensorlayer} is designed for researchers to customize neural networks for various applications. TensorLayer is a wrapper of TensorFlow and supports the OpenAI gym-style environments. However, it is not user-friendly.

%\item \textbf{Garage} \cite{garage} features comprehensive automated unit test and benchmarking suite it benchmarks the full suite of algorithms to detect regressions and improvements that may have missed. 

%\item \textbf{RLlab} \cite{duan2016benchmarking} is a DRL framework for developing continuous control tasks. 

%\item \textbf{Tensorforce} \cite{lift-tensorforce} is an extension of TensorFlow and preserves its benefits. However, the codes are complicated, making it not friendly to beginners.

%\item \textbf{Horizon} \cite{gauci2018horizon} is a deep learning framework using PyTorch, whose main feature is to train DRL agents in a batch setting. However, there are no good support for a distributed or scalable infrastructure.

%{\color{red}However, such libraries.....lack of financial support}.

%%%%%%%%%%%%%%%%%%%%%%%%%%%%%%%%%%%%%%%%%%%%%%%%%%%
\subsection{Deep Reinforcement Learning in Finance}

%\yanglet{There are many more works using DRL. Check the ICAIF 2020 website: \url{https://ai-finance.org/conference-program/} and also previous finance workshop at NeurIPS 2018, 2019, ICML 2019, KDD 2019, etc. Find the names of those workshop at (\url{http://www.tensorlet.com/projects/ai-in-finance/}) and Google them to direct to the homepage.}

Many recent works have applied DRL to quantitative finance. Stock trading is considered as the most challenging task due to its noisy and volatile features, and various DRL based approaches \cite{ Nan2020SentimentAK, zhang2020deep, Ganesh2018DeepRL} have been applied. Volatility scaling was incorporated in DRL algorithms to trade futures contracts, which considered market volatility in a reward function. News headline sentiments and knowledge graphs, as alternative data, can be combined with the price-volume data as time series to train a DRL trading agent. High frequency trading using DRL \cite{HFT_Rundo_2019} is a hot topic. 

Deep Hedging \cite{buehler2019deep} designed hedging strategies with DRL algorithms that manages the risk of liquid derivatives. It has shown two advantages of DRL in mathematical finance, \textit{scalable} and \textit{model-free}. DRL driven strategy would become more efficient as the scale of the portfolio grows. It uses DRL to manage the risk of liquid derivatives, which indicates further extension of our FinRL library into other asset classes and topics in mathematical finance. 

%, but also decouples from the choice of a financial market simulator. 
%It uses DRL to manage the risk of liquid derivatives, which indicates further extension of our library into other asset classes and topics in mathematical finance. 

Cryptocurrencies are rising in the digital financial market, such as Bitcoin (BTC) \cite{Sapuric2014BitcoinIV}, and are considered more volatile than stocks. DRL is also being actively explored in automated trading, portfolio allocation, and market making for cryptocurrencies \cite{BTC_2020_Sattarov,Jiang2017CryptocurrencyPM, Sadighian2019DeepRL}.
%The open-source Python ecosystem such as NumPy \cite{harris2020array}, SciPy \cite{2020SciPy-NMeth}, Pandas \cite{mckinney-proc-scipy-2010}, Scikit-learn \cite{scikit-learn}, Keras \cite{chollet2015keras}, TensorFlow \cite{tensorflow2015-whitepaper}, Pytorch \cite{NEURIPS2019_9015}, OpenAI \cite{brockman2016openai} are widely used by nearly all the world’s major AI companies. This fulfilled needs of statistical analysis, scientific computing, machine learning application, deep learning and reinforcement learning research in the AI community. 

% \section3{The Proposed FinRL Library}
%\bruce{In the following, we should emphasize one or two feature of each layer. Currently, it focuses on description, while we need to sell it.}

\section{The Proposed FinRL Framework}
%第一阶段，你这边把想法写下来，以完整为目标，把逻辑尽可能捋顺；  第二阶段，我们内部做一次讨论，开始修补一些点，争取思考得周全，兼顾，完善，有说服力；  第三阶段： 裁剪，美化成会议的篇幅。
%为什么要支持这个库，开源，然后把credits给你？
%逻辑上，强化学习为什么被看好
%便捷，省了dirty work，做到界面类的，商业化
%对业界的人的benifit和对学术界的人的benifit
%既然有tensorflow和pytorch深度学习，为什么要开发finrl？TF和Pytorch的人能够迅速开发出这样一个库
%ElegantRL核心部件，做金融和做普通的强化学习的库，有没有解决特定的问题？为Finance做的特定的事儿？针对金融？

%RL本身跟ML的区别；Finance的application跟robotics和control不一样；为什么Finance跟他们的东西不一样要单独开发一个toolbox。环境互动的引擎速度慢，ML是input和output是available的。游戏的话要施加到游戏的引擎上，如果是时间序列的话可以直接查的，Finance的RL离robotics远，离ML近。采样的过程影响整个RL速度。
%快速训练：数据本身的特点，拿到别的问题解，
%advantage是什么：
%问题的本质不一样，已有的工具的困难，我们可以解决困难：耗时上面和性能上面
%benchmark，metrics去衡量训练耗时，stable-baseline v.s. elegantRL，高频交易时间
%gym变成业界的标准
%说服力提上去
%分类：基于historical data (offline RL) 和 Live Trading API (Online RL)
%tick data市价和limit order book限价
%ccxt: online and offline
%为什么要用histroical和live？用live data(production mode去评估模型的真实效果)去评估模型的好坏，这样结果能够让人更信服，所以我们的衡量指标更准确。
%DB怎么跟RL构建关联的？
%为了支持三个算法，我做了什么东西？

% The FinRL library available research, papers, and open-source repositories to get a better understanding of deep reinforcement learning and trading.
%Our proposed FinRL library consists of applications including Stock Trading, Portfolio Allocation, Liquidation Strategy, Benchmark Test and User-defined Trading Tasks, as shown in the “Application Layer” in Fig.1. The library is designed with a scalable and accessible fashion. 

%FinRL has a three-layer architecture, as shown in Fig. \ref{overview_finrl}.

In this section, we first present an overview of the FinRL framework and describe its layers. Then, we propose a training-testing-trading pipeline as a standard evaluation of the trading performance.

\subsection{Overview of FinRL Framework}

%All the functions are divided into the several modules, therefore, the applications of FinRL can be invoked through terminal commands:\\
%python -m baselines.run --alg=\underline{\emph{algorithm's name}} \\--env=\underline{\emph{environment's name}} --network=mlp \\--num\_timesteps=\underline{\emph{timesteps}},\\
%and the commands include importing training environment, invoking DRL agent, deploying network and outputting test results.
% and provides respective layers with standard APIs, which 

%FinRL library has a layered architecture that separates learning agents and the supporting applications, thus making it agnostic to the type of interaction with the application environment. 

%Layered architecture不是特性
%Modularity是为了达到flexibility，增加灵活性，
%transparency，对用户透明，以前要管数据格式是什么，减少用户的effort，用户不用管下面的API，底层数据库调用透明，降低学习effort。所有都是我们底层帮你做的。抽象层。adapt适配。什么的抽象层。核心contribution，脏活累活
%histroical和live的区别：历史数据是人工处理过的，改动了，不能真实反映当时真实情况。live数据是完全准确的数据。
%live是真实的环境，生成结果的。live这个API是个env
%用历史数据集和live数据集的区别和关联。
%RL策略子学习，

The FinRL framework has three layers, application layer, agent layer, and environment layer, as shown in Fig. \ref{overview_finrl}. %constituent stocks from NASDAQ-100, DJIA, S\&P 500, SSE 50, CSI 300, HSI are supported.  %FinRL builds various financial environments for training, depending on historical data of constituents of NASDAQ-100, DJIA, S\&P 500, SSE 50, CSI 300, HSI. This layer defines the action spaces (buying, holding, selling) and the state spaces (shares of stocks, remaining balance, price of stocks).On the agent layer, FinRL calls and fine-tunes DRL algorithms from existing DRL libraries (in a plug-and-play manner). Upper-layer trading tasks can directly call DRL algorithms, e.g., DQN \cite{mnih2015human}, DDPG \cite{lillicrap2015continuous}, multi-agent DDPG \cite{lowe2017multi}, PPO \cite{schulman2017proximal}, SAC \cite{haarnoja2018soft}, A2C \cite{mnih2016asynchronous} and TD3 \cite{fujimoto2018addressing}.
\begin{itemize}[leftmargin=*]
    \item On the application layer, FinRL aims to provide hundreds of demonstrative trading tasks, serving as stepping stones for users to develop their strategies.
    \item On the agent layer, FinRL supports fine-tuned DRL algorithms from DRL libraries in a plug-and-play manner, following the unified workflow in Fig. \ref{state_transition}.
    \item On the environment layer, FinRL aims to wrap historical data and live trading APIs of hundreds of markets into training environments, following the defacto standard Gym \cite{brockman2016openai}.
\end{itemize}
 Upper-layer trading tasks can directly call DRL algorithms in the agent layer and market environments in the environment layer. 

The FinRL framework has the following features:
\begin{itemize}[leftmargin=*]
    \item \textbf{Layered architecture}: The lower layer provides APIs for the upper layer, ensuring \textit{transparency}. The agent layer interacts with the environment layer in an exploration-exploitation manner. Updates in each layer is independent, as long as keeping the APIs in Table \ref{tab:FinRL_API} unchanged.
    %making the lower layer transparent to the upper layer. %The agent layer interacts with the environment layer in an exploration-exploitation manner.
    %There are three layers: trading applications, trading agents and market environments.  
    \item \textbf{Modularity and extensibility}: Each layer has modules that define self-contained functions. A user can select certain modules to implement her trading task. We reserve interfaces for users to develop new modules, e.g., adding new DRL algorithms.
    \item \textbf{Simplicity and applicability}: FinRL provides benchmark trading tasks that are reproducible for users, and also enables users to customize trading tasks via simple configurations. In addition, hands-on tutorials are provided in a beginner-friendly fashion. 
\end{itemize}

% develop their own strategies.

\subsection{Application Layer}

On the application layer, users map an algorithmic trading strategy into the DRL language by specifying the state space, action space and reward function. For example, the state, action and reward for several use cases are given in Table \ref{event:eventTypes}. Users can customize according to their own trading strategies.

%The state space representing the environment is used to pick the action in the action state and the reward as incentive. Every trading day, price of stocks in the state space changes and the agent takes trading actions on this basis.  

%It serves with supervision under real market and is more realistic and robust compared to artificial environment for DRL.
%why real market
%data set 
%\vspace{-0.05in}
%\subsubsection{\textbf{State Space, Action Space, and Reward Function}} 

\begin{table}[t]
\small
\renewcommand{\arraystretch}{1.2}
\centering
\begin{tabular}{|l|l|m{200pt}<{\centering}|}
   \hline   
    \textbf{Key components} & \textbf{Attributes} \\
   \hline
    \multirow{4}{2cm}{State} & Balance ${b}_{t}\in \mathbb{R}_+$;~~~~Shares $\bm{k}_{t}\in \mathbb{Z}_+^{n}$ \\
    %\arrayrulecolor{gray} \cdashline{2-3}[0.8pt/2pt]
    %& Shares $\bm{h}_{t}\in \mathbb{Z}_+^{n}$  \\
    %\cdashline{2-3}[0.8pt/2pt]
    %& Closing price $\bm{p}_{t}\in \mathbb{R}_+^{n}$  \\
    \cdashline{2-3}[0.8pt/2pt]
    & OHLCV data $\bm{o}_{t}, \bm{h}_{t}, \bm{l}_{t},\bm{p}_{t},\bm{v}_{t} \in \mathbb{R}_+^{n}$ \\
    \cdashline{2-3}[0.8pt/2pt]
    & Technical indicators; Fundamental indicators \\
    \cdashline{2-3}[0.8pt/2pt]
    & Smart beta \\
    \cdashline{2-3}[0.8pt/2pt]
    & NLP market sentiment features\\
    %\cdashline{2-3}[0.8pt/2pt]
    %& Fundamental indicators \\
    \hline
     \multirow{2}{2cm}{Action} & Buy/Sell/Hold;~~~~~~Short/Long   \\
    %\cdashline{2-3}[0.8pt/2pt]
    %& Short/Long \\
    \cdashline{2-3}[0.8pt/2pt]
    & Portfolio weights \\
    \hline
     \multirow{3}{2cm}{Rewards} & Change of portfolio value \\  
    \cdashline{2-3}[0.8pt/2pt]
    & Portfolio log-return \\
    \cdashline{2-3}[0.8pt/2pt]
    & Shape ratio \\  
    \hline
     \multirow{3}{2cm}{Environment} &  Dow-$30$, NASDAQ-$100$, S\&P-$500$ \\  
    % \cdashline{2-3}[0.8pt/2pt]
    % & NASDQQ-$100$ \\
    \cdashline{2-3}[0.8pt/2pt]
    & Cryptocurrencies \\ 
    \cdashline{2-3}[0.8pt/2pt]
    & Foreign currency and exchange \\ 
    \cdashline{2-3}[0.8pt/2pt]
    & Futures and options \\ 
    \cdashline{2-3}[0.8pt/2pt]
    & Living trading \\
   \hline
\end{tabular}
\caption{Key components and attributes. OHLCV stands for Open, High, Low, Close and Volume.}\vspace{-0.25in}
\label{event:eventTypes}
\vspace{-0.1in}
\end{table}
%重画Fig2
%做个各个space的表格
%我们的environment包括两个部分，前两个是自己的环境，第三个是market（在文中表述清楚}

\textbf{State space $\mathcal{S}$}.
The state space describes how the agent perceives the environment. A trading agent observes many features to make sequential decisions in an interactive market environment. We allow the time step $t$ to have \textit{multiple levels of granularity}, e.g., daily, hourly or a minute basis. We provide various features for users to select and update, in each time step $t$:
\begin{itemize}[leftmargin=*]
\item Balance ${b}_{t}\in \mathbb{R}_+$: the account balance at the current time step $t$.
\item Shares $\bm{k}_{t}\in \mathbb{Z}_+^{n}$: current shares for each asset, where $n$ represents the number of stocks in the portfolio.
%\item Close price $\bm{p}_{t}\in \mathbb{R}_+^{n}$: one commonly used feature.
\item Open-high-low-close (OHLC) prices $\bm{o}_{t}, \bm{h}_{t}, \bm{l}_{t}, \bm{p}_{t} \in \mathbb{R}_+^{n}$ and trading volume $\bm{v}_{t} \in \mathbb{R}_+^{n}$.
%\item Trading volume $\bm{v}_{t}\in \mathbb{R}_+^{n}$: total quantity of shares traded during a trading slot.
\item Technical indicators, including Moving Average Convergence Divergence (MACD) $\bm{M}_t \in \mathbb{R}^{n}$, Relative Strength Index (RSI) $\bm{R}_t \in \mathbb{R}_+^{n}$, etc.
\item Fundamental indicators, including return on assets (ROA), return on equity (ROE), net profit margin (NPM), price-to-earnings (PE) ratio, price-to-book (PB) ratio, etc.
\end{itemize}

\textbf{Action space $\mathcal{A}$}. 
The action space describes the allowed actions that an agent can take at a state. An action of one share is $a \in \{-1, 0, 1\}$ where $-1, 0, 1$ represent selling, holding, and buying, respectively; an action of multiple shares is $a \in \{-k,...,-1, 0, 1, ..., k\}$ where $k$ denotes the maximum number of shares to buy or sell, e.g., "Buy/Sell 10 shares of AAPL" is $10$ or $-10$, respectively. 

%The continuous action space needs to be normalized to $[-1, 1]$, since the policy is defined on a Gaussian distribution, which needs to be normalized and symmetric.

\textbf{Reward function}.
The reward function $r(s,a,s')$ is the incentive for an agent to learn a better policy. FinRL supports user-defined reward functions to reflect risk-aversion or market friction \cite{zhang2020deep,buehler2019deep} and provides commonly used ones \cite{RL_survey} as follows:
\begin{itemize}[leftmargin=*]
\item The change of the portfolio value when taking action $a$ at state $s$ and arriving at new state $s'$ \cite{xiong2018practical,Nan2020SentimentAK, yang2020},
$r(s,a,s') =  v' - v$, where $v'$ and $v$ are portfolio values at $s'$ and $s$, respectively.
\item The portfolio log return \cite{huang2018financial}, $ r(s,a,s') =  \log(v' / v)$.
\item The Sharpe ratio for trading periods $t=1,...,T$ \cite{Moody1998PerformanceFA},
\begin{equation} \label{reward3}
    \text{Sharpe ratio} =(\mathbb{E}\left(R_{t}\right)-R_{f}) / \text{std}(R_t),
\end{equation}
where $R_t =  v_t - v_{t-1}$, and $R_f$ is the risk-free rate.
\end{itemize}

%\textbf{Action-value function $Q_{\pi}(s, a)$}.
%Action-value function $Q_{\pi}(s, a)$ means the expected value of taking action $a$ in state $s$ under policy $\pi$.

%%%%%%%%%%%%%%%%%%%%%%%%%%%%%%%%%%%%%%%%%%%%%%%%%%%%%%%%%%%%%%%%%%%%%%%%%%%%%%%%%%
\subsection{Agent Layer}

%第一部分展示完整的数据接口和支持
%第二部分接通和支持三种DRL库
%第三部分支持比较完善的backtesting给于样例

%The DRL algorithms under the Actor-Critic framework, where an agent consists of an Actor network and a Critic network. Due to the completeness and simplicity of code structure, users are able to easily customize their own agents.

FinRL allows users to plug in and play with standard DRL algorithms, following the unified workflow in Fig. \ref{state_transition}. As a backbone, we fine-tune three representative open-source DRL libraries, namely Stable Baseline 3 \cite{stable-baselines}, RLlib \cite{liang2018rllib} and ElegantRL \cite{erl}. User can also design new DRL algorithms by adapting existing ones.

\subsubsection{Agent APIs}
%中间层，链接env，输出结果
%统一调用
%input: env，不同market的env
%output: actions or trading positions

FinRL uses unified Python APIs for training a trading agent. The Python APIs are flexible so that a DRL algorithm can be easily plugged in. To train a DRL trading agent, as in Fig. \ref{overview_finrl}, a user chooses an environment (i,e., StockTradingEnv, StockPortfolioEnv) built on historical data or live trading APIs with default parameters (env\_kwargs), and picks a DRL algorithm (e.g., PPO \cite{schulman2017proximal}). Then, FinRL initializes the agent class with the environment, sets a DRL algorithm with its default hyperparameters (model\_kwargs), then launches a training process and returns a trained model.

The main APIs are given in Table \ref{tab:FinRL_API}, while the details of building an environments, importing an algorithm, and constructing an agents are hidden in the API calls. 

\begin{table*}
\begin{center}
\begin{tabular}{|p{7.5cm}|p{9.5cm}|}
\hline
\textbf{Function} &\textbf{Description}\\
\hline
env = StockTradingEnv(df, **env\_kwargs)
& Return an environment instance of the Env class with data and default parameters.\\
\hline
agent = DRLAgent(env) 
& Instantiate a DRL agent with a given environment env.\\
\hline
model = agent.get\_model(model\_name, **model\_kwargs)
& Return a model with name and default hyperparameters.\\
\hline
trained\_model = agent.train\_model(model)
& Launch the training process for the agent and return a trained model.\\
\hline
\end{tabular}
\end{center}
\caption{Main APIs of FinRL.}\vspace{-0.2in}
\label{tab:FinRL_API} 
\vspace{-0.1in}
\end{table*}

%%%%%%%%%%%%%%%%%%%%%%%%%%%%%%%%%%%%%%%%%%%%%%%%%%%%%%%%%%%%%
\subsubsection{Plug-and-Play DRL Libraries}

\begin{figure}
 \centering
 \includegraphics[height=3.5cm]{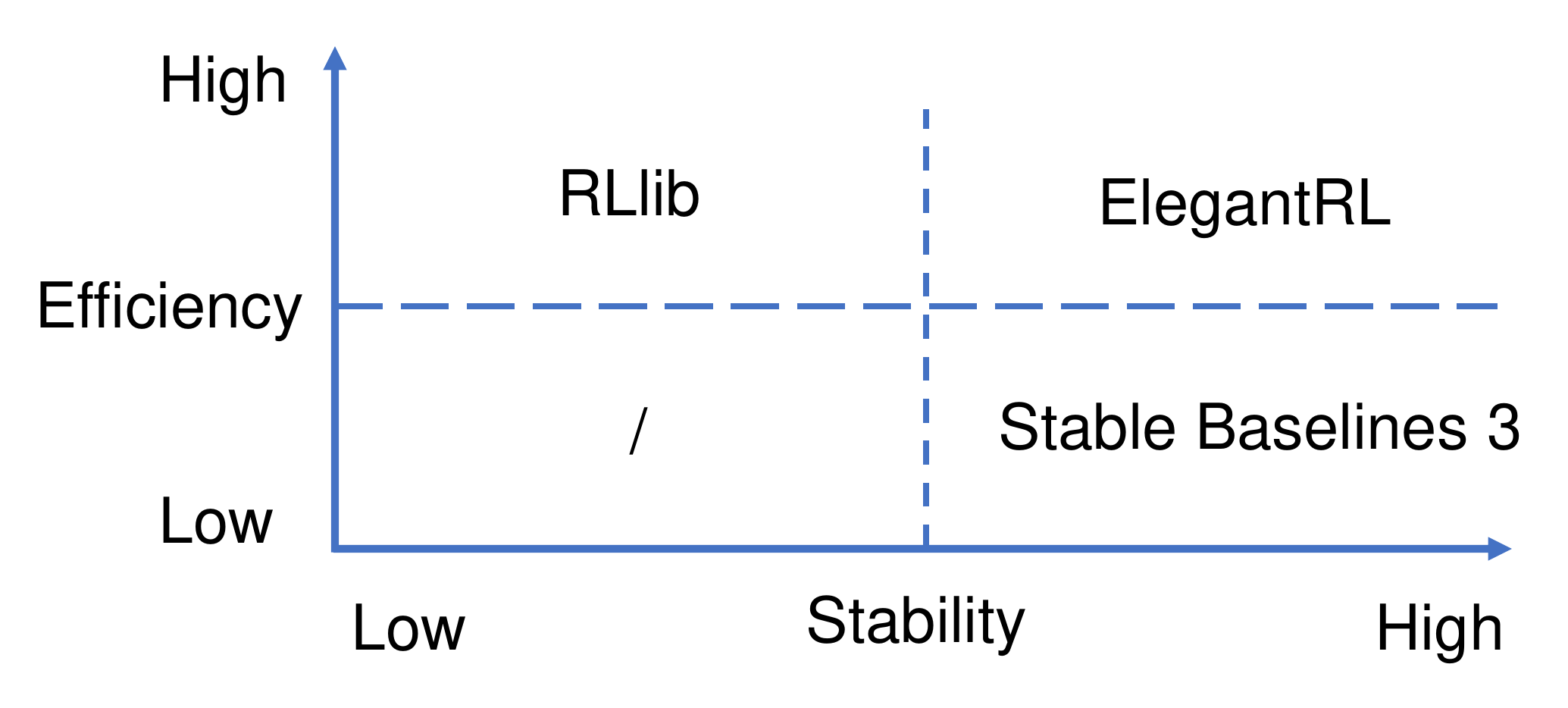}\vspace{-0.2in}
 \caption{Comparison of DRL libraries.}\vspace{-0.2in}
 \label{RLlab}
 \vspace{-0.1in}
 \end{figure}

Fig. \ref{RLlab} compares the three DRL libraries. The details of each library are summarised as follows.

\textbf{Stable Baselines 3} \cite{stable-baselines} is a set of improved implementations of DRL algorithms over the OpenAI Baselines \cite{baselines}. FinRL chooses to support Stable baselines 3 due to its \textbf{advantages}: 1). User-friendly, 2). Easy to replicate, refine, and identify new ideas, and 3). Good documentation.  Stable Baselines 3 is used as a base around which new ideas can be added, and as a tool for comparing a new approach against existing ones. The purpose is that the simplicity of these tools will allow beginners to experiment with a more advanced tool set, without being buried in implementation details. 

%(user friendly, widely used, etc)
%\subsubsection{RLlib}

\textbf{RLlib} \cite{liang2018rllib} is an open-source high performance library for a variety of general applications. FinRL chooses to support RLlib due to its \textbf{advantages}: 1). High performance and parallel DRL training framework; 2). Scale training onto large-scale distributed servers; and 3). Allowing the multi-processing technique to efficiently train on laptops. RLlib natively supports TensorFlow, TensorFlow Eager, and PyTorch, but most of its internals are framework agnostic.

%(excellent performance and efficiency, good for servers use)

\textbf{ElegantRL} \cite{erl} is designed for researchers and practitioners with finance-oriented optimizations. FinRL chooses to support ElegantRL due to its \textbf{advantages}: 1). Lightweight: core codes have less than 1,000 lines, less dependable packages, only using PyTorch (train), OpenAI Gym \cite{brockman2016openai} (env), NumPy, Matplotlib (plot);  2). Customization: Due to the completeness and simplicity of the code structure, users can easily customize their own agents; 3). Efficient: Performance is comparable with RLlib \cite{liang2018rllib}; and 4). Stable:  As stable as Stable baseline 3 \cite{stable-baselines}.

ElegantRL supports state-of-the-art DRL algorithms, including both discrete and continuous ones, and provides user-friendly tutorials in Jupyter Notebooks. ElegantRL implements DRL algorithms under the Actor-Critic framework, where an agent consists of an actor network and a critic network. The ElegantRL library enables researchers and practitioners to pipeline the disruptive “design, development and deployment” of DRL technology.

\textbf{Customizing trading strategies}. Due to the uniqueness of different financial markets, customization becomes a vital character to design trading strategies. Users are able to select a DRL algorithm and easily customize it for their trading tasks by specifying the state-action-reward tuple in Table \ref{event:eventTypes}. We believe that among the three state-of-the-art DRL libraries, \textbf{ElegentRL} is a practically useful option for financial tasks because of its completeness and simplicity along with its comparable performance with RLlib \cite{liang2018rllib} and stability with Stable Baselines 3 \cite{stable-baselines}.

%(best for Customization, also absorb advantage of 1 and 2)

%\subsection{Backtesting Examples}

%\yanglet{Data structure: more efficient}

%\yanglet{Feature; Technical Indicators}

%\yanglet{High frequency trading (Multi-GPUs)}

\vspace{-0.2in}
\subsection{Environment Layer}

%env在framework里是一个什么位置，对于外行人讲是一个什么，输入是什么输出是什么。具体解释为什么要这么做。MDP问题如何转换，framework。
%过渡，一个env在你这个framework里面是干什么的，在乎的是接收数据的，看当前状态的。
Environment design is crucial in DRL, because the agent learns by interacting with the environment in a trial and error manner. A good environment that simulates real-world market will help the agent learn a better strategy. Considering the stochastic and interactive nature, a financial task is modeled as a Markov Decision Process (MDP), whose state transition is shown in Fig. \ref{state_transition}. 

The environment layer in FinRL is responsible for observing current market information and translating those information into states of the MDP problem. The state variables can be categorized into the state of an agent and the state of the market. For example, in the use case stock trading, the state of the market includes the open-high-low-close prices and volume (OHLCV) and technical indicators; the state of an agent includes the account balance and the shares for each stock.

The RL training process involves observing price change, taking an action and calculating a reward. By interacting with the environment, the agent updates iteratively and eventually obtains a trading strategy to maximize the expected return. We reconfigure real market data into gym-style training environments according to the principle of \textit{time-driven simulation}. Inspired by OpenAI Gym \cite{brockman2016openai}, FinRL provides strategy builders with a collection of universal training environments for various trading tasks.
\vspace{-0.1in}
\subsubsection{Standard Datasets and Live Trading APIs}

DRL in finance is different from chess, card games and robotics \cite{silver2016mastering,Zha2019ExperienceRO}, which may have physical engines or simulators. Different financial tasks may require different market simulators. Building such training environments is time-consuming, so FinRL provides a set of representative ones and also supports user-import data, aiming to free users from such tedious and time-consuming work.
 
\noindent\textbf{NASDAQ-100 index constituents} are $100$ stocks that are characterized by high technology and high growth. %Training in such an environment gets agent to capture the financial trends of technology.

\noindent\textbf{Dow Jones Industrial Average (DJIA) index} is made up of $30$ representative constituent stocks. DJIA is the most cited market indicator to examine market overall performance. %It is made up of 30 constituents that best represent their industries respectively. %Many researches select DJIA as an indicator of overall market performance for their agents.

\noindent\textbf{Standard \& Poor's 500 (S\&P 500) index constituents} consist of $500$ largest U.S. publicly traded companies.

\noindent\textbf{Hang Seng Index Index (HSI) constituents} are grouped into Finance, Utilities, Properties and Commerce \& Industry \cite{HSI}. HSI is the most widely quoted indicator of the Hong Kong stock market. 
%HSI constituent securities are grouped into Finance, Utilities, Properties and Commerce \& Industry \cite{HSI}. 

\noindent\textbf{SSE 50 Index constituents} \cite{SSE} include the best representative companies (in $10$ industries) of A shares listed at Shanghai Stock Exchange (SSE) with considerable size and liquidity. %Listed companies are classified into 10 industries \cite{SSE}. 

\noindent\textbf{CSI 300 Index constituents} \cite{CSI300} consist of the $300$ largest and most liquid A-share stocks listed on Shenzhen Stock Exchange and SSE. This index reflects the performance of the China A-share market.

\noindent\textbf{Bitcoin (BTC) Price Index} consists of the quote and trade data on Bitcoin market, available at \url{https://public.bitmex.com/}.

\vspace{-0.1in}
\subsubsection{User-Imported Data}
Users may want to train agents on their own data sets. FinRL provides convenient support for users to import data, adjust the time granularity, and perform the training-testing-trading data split. We specify the format for different trading tasks, and users preprocess and format the data according to our instructions. Stock statistics and indicators can be calculated using our support, which provides more features for the state space. Furthermore, episodic total return and Sharpe ratio can also assist performance evaluation.
% User-defined Environment

%\textbf{Training phase}: we use data from 01/01/2009 to 09/30/2015 to train the automated trading agents.

%\textbf{Testing phase}: we use data from 10/01/2015 to 12/31/2015 to validate the results and adjust key parameters, such as time slice's size and learning rate.

%\textbf{Trading phase}: we use data from 01/01/2016 to 05/08/2020 (unseen data) to evaluate the performance of the applications.

%In performance evaluation of RL algorithms, the portfolio value and stock index yield are compared in the form of curve graphs in fig.3), users can see the performance of the algorithm on the DJIA constituents' data. Then specific indicators are shown in Table 1, showing the risks of trading using algorithms.

%From the results of RL, we can conclude that DRL method have better performance on DJIA than traditional investment methods and RL methods. There are also differences among different DRL methods. The evaluation model of FinRL is multi-angled and effective.

%\begin{figure*}[t]
%\centering
%\includegraphics[height=3.8cm]{image/Training-Testing-Trading.png}
%\caption{Overview of the training-testing-trading pipeline.}
%\label{data_flow}
%\vspace{-0.15in}
%\end{figure*}

%\vspace{-0.10in}
%%%%%%%%%%%%%%%%%%%%%%%%%%%%%%%%%%%%%%
\vspace{-0.1in}
\subsection{Training-Testing-Trading Pipeline}

\begin{figure}[t]
\centering
\includegraphics[height=2.9cm]{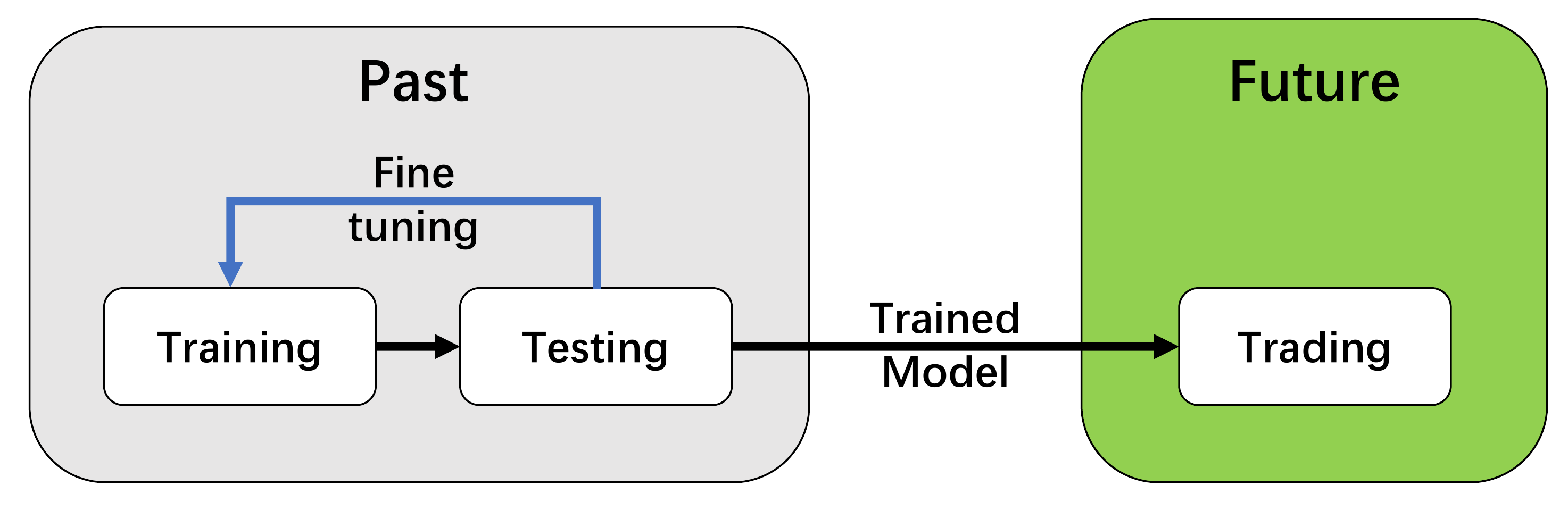}\vspace{-0.2in}
\caption{The training-testing-trading pipeline.}\vspace{-0.1in}
\label{data_flow}
\vspace{-0.1in}
\end{figure}

%\yanglet{1. Why the "training-testing" workflow in machine learning fall short for financial tasks?}

%\yanglet{To Bruce: need to give reasons about why this flow is necessary for financial tasks?}
The "training-testing" workflow used by conventional machine learning methods falls short for financial tasks. It splits the data into training set and testing set. On the training data, users select features and tune parameters; then evaluate on the testing data. However, financial tasks will experience a \textit{simulation-to-reality gap} between the testing performance and real-live market performance. Because  the testing here is offline backtesting, while the users' goal is to place orders in a real-world market. 

FinRL employs a ``training-testing-trading" pipeline to reduce the simulation-to-reality gap. We use historical data (time series) for the ``training-testing" part, which is the same as conventional machine learning tasks, and this testing period is for backtesting purpose. For the ``trading" part, we use live trading APIs, such as CCXT, Alpaca, or Interactive Broker, allowing users carry out trades directly in a trading system. Therefore, FinRL directly connects with live trading APIs: 1). downloads live data, 2). feeds data to the trained DRL model and obtains the trading positions, and 3). allows users to place trades.

%\begin{figure*}[t]
%\centering
%\includegraphics[height=2.0cm]{image/timeline507.png}
%\caption{Data splitting.}
%\label{fig4}
%\end{figure*}

Fig. \ref{data_flow} illustrates the “training-testing-trading” pipeline:

\noindent \textbf{Step 1)}. A training window to retrain an agent. 

\noindent \textbf{Step 2)}. A testing window to evaluate the trained agent, while hyperparameters can be tuned iteratively.

\noindent \textbf{Step 3)}. Use the trained agent to trade in a trading window.

Rolling window is used in the training-testing-trading pipeline, because the investors and portfolio managers need to retrain the model periodically as time goes ahead. FinRL provides flexible selections of rolling windows, such as monthly, quarterly, yearly windows, or by users' specifications.

% section4: Performance Evaluation
%\input{sections/Section5-PerformanceEvaluation}

% section5: Demo Applications 
%\begin{figure*}
%\centering
%\includegraphics[width=1\textwidth]{image/perform_1_1.png}
%\caption{Performance of stock trading %\cite{yang2020} using the FinRL library.}
%\label{fig3}
%\end{figure*}
\begin{figure*}
\centering
\includegraphics[width=0.95\textwidth]{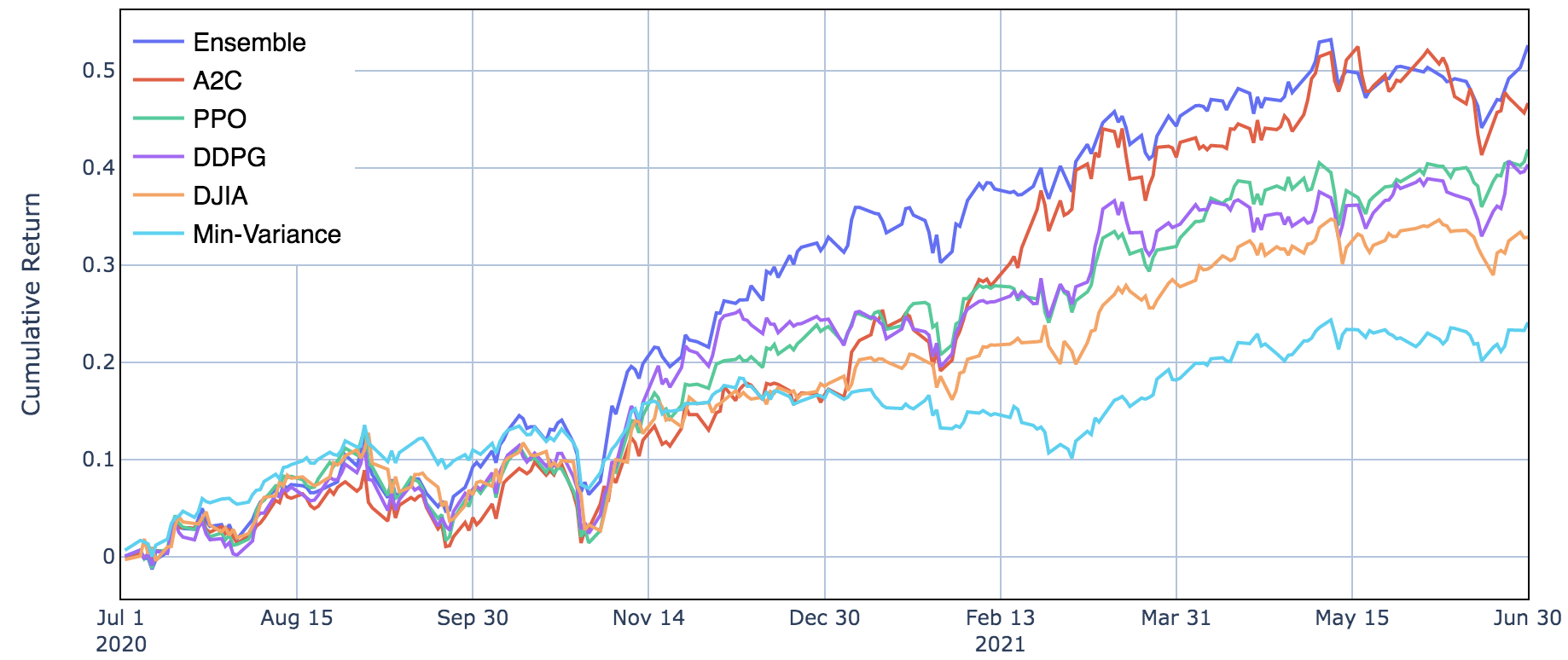}\vspace{-0.2in}
\caption{Performance of stock trading \cite{yang2020} using the FinRL framework.}
\label{stock_trading_1}
\vspace{-0.1in}
\end{figure*}

\begin{figure*}
\centering
\includegraphics[width=0.95\textwidth]{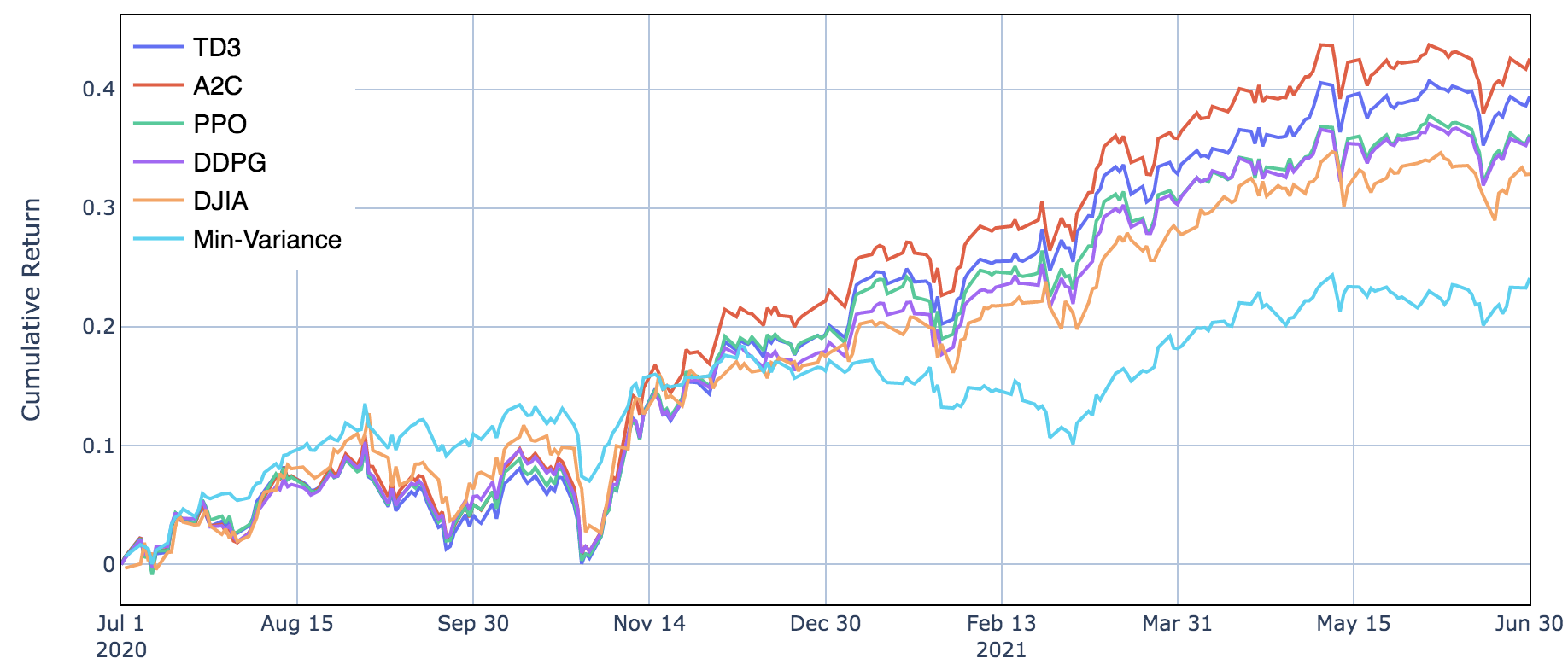}\vspace{-0.2in}
\caption{Performance of portfolio allocation \cite{Jiang2017ADR} using the FinRL framework.}
\label{portfolio_1}
\vspace{-0.1in}
\end{figure*}

\begin{table*}
		\centering
		%\resizebox{\textwidth}{!}{
		\begin{tabular}{|l|c|c|c|c|c|c|c|c|}\hline
			07/01/2020-06/30/2021 & Ensemble \cite{yang2020} & A2C & PPO&  DDPG& TD3 & Min-Var. &  DJIA \\
			\hline
			Initial value &1M&1M &1M&1M&1M&1M&1M\\
			Final value &\textcolor{red}{1.52M}& \textcolor{red}{1.46M};~~\textcolor{blue}{1.43M}
			& \textcolor{red}{1.42M};~~\textcolor{blue}{1.36M}
			& \textcolor{red}{1.40M};~~\textcolor{blue}{1.36M}
			&\textcolor{blue}{1.39M}
			&1.24M&1.33M\\
			
			Annualized return & \textcolor{red}{52.61\%}
			&\textcolor{red}{46.65\%}; ~~\textcolor{blue}{42.57\%}
			&\textcolor{red}{41.90\%}; ~~\textcolor{blue}{36.17\%}
			&\textcolor{red}{40.34\%}; ~~\textcolor{blue}{36.01\%}
			&\textcolor{blue}{39.38\%}
			&24.10\%& 32.84\%\\
			Annualized Std &\textcolor{red}{15.53\%}
			&\textcolor{red}{17.86\%}; ~~\textcolor{blue}{15.51\%}
			&\textcolor{red}{16.33\%};~~\textcolor{blue}{15.20\%}
			&\textcolor{red}{17.28\%};~~\textcolor{blue}{14.38\%}
			&\textcolor{blue}{15.08\%}
			&11.2\%&14.5\%\\
			Sharpe ratio     
			&\textcolor{red}{2.81}
			& \textcolor{red}{2.24}; ~~\textcolor{blue}{2.36}
			&\textcolor{red}{2.23}; ~~\textcolor{blue}{2.11} 
			&\textcolor{red}{2.05}; ~~\textcolor{blue}{2.21} 
			&\textcolor{blue}{2.28} 
			& 1.98 &2.02  \\
			Max drawdown 
			&\textcolor{red}{-7.09\%}
			&\textcolor{red}{-7.59\%};~~\textcolor{blue}{-9.04\%}&\textcolor{red}{-9.41\%}; ~~\textcolor{blue}{-8.68\%}
			&\textcolor{red}{-8.10\%}; ~~\textcolor{blue}{-8.46\%}
			&\textcolor{blue}{-8.92\%}
			&-6.97\%&-8.93\%\\
			\hline
		\end{tabular}
		\caption{Performance of \textcolor{red}{stock trading} and \textcolor{blue}{portfolio allocation} over the DJIA constituents stocks using FinRL. The Sharpe ratios of the ensemble strategy and the individual DRL agents excceed those of the DJIA index, and of the min-variance strategy.}\vspace{-0.2in}
		\label{tab:Performance evaluation}
		\vspace{-0.1in}
	\end{table*}
\vspace{-0.1in}
\section{Hands-on Tutorials and Benchmark Performance}

We provide hands-on tutorials and reproduce existing works as use cases. Their configurations and commands are available on Github. 

%Use case复现论文。
%Tutorial 在手动教步骤。

%\subsection{Performance Metrics}
%\label{sec:performance_metrics}

%FinRL includes common metrics to evaluate trading performance.

%\begin{equation}
%\text{Sharpe ratio} =(\mathbb{E}\left(R_{p}\right)-R_{f}) / \sigma_{p},
%\end{equation}
%where $\mathbb{E}\left(R_{p}\right)$ is expected return rate, $R_{f}$ is risk-free rate, and $\sigma_{p}$ is the standard deviation of $R_{p}$. 

%%%%%%%%%%%%%%%%%%%%%%%%%%%%%%%%%%%%%%
\vspace{-0.1in}
\subsection{Backtesting Module}
\label{sec:backtesting_module}

Backtesting plays a key role in evaluating a trading strategy. FinRL library provides an automated backtesting module based on Quantopian pyfolio package \cite{pyfolio}. It is easy to use and consists of various individual plots that provide a comprehensive image of the performance. In order to facilitate users, FinRL also incorporates market frictions, market liquidity and the investor's degree of risk-aversion. %All of those are crucial among key determinants of net returns. 

%In order to facilitate users, we also incorporate trading constraints and risk-aversion in the backtesting module. builds a collection of market environments in OpenAI Gym style \cite{brockman2016openai}, and 

\vspace{-0.1in}
\subsubsection{Incorporating Trading Constraints}

Transaction costs incur when executing a trade, such as broker commissions and the SEC fee. We allow users to treat transaction costs as parameters in the environments: 1). \textbf{Flat fee} is a fixed amount per trade; and 2). \textbf{Per share percentage} is a percentage rate for every share, e.g., $0.1 \%$ or $0.2 \%$ are most commonly used.

Moreover, we need to consider market liquidity for stock trading, e.g., the bid-ask spread that is the difference between the best bid and ask prices. In our environment, users can add the bid-ask spread as a parameter. For different levels of risk-aversion, 
users can add the standard deviation of the portfolio returns into the reward function or use a risk-adjusted Sharpe ratio as the reward function.

\subsubsection{Risk-aversion}
An investor may prefer conservative trading in highly volatile markets. For a worst case scenario as the $2008$ global financial crisis, FinRL employs the turbulence index $\text{turbulence}_t$ to measure extreme fluctuation \cite{turbulence}:
\begin{equation}  
\text{turbulence}_t = \left(\bm{y_t} - \bm{\mu}\right)^{T}\bm{\Sigma^{-1}}(\bm{y_t}-\bm{\mu}) \in \mathbb{R},
\label{turb}
\end{equation} 
where $\bm{y_t} \in \mathbb{R}^n$ is the return at $t$, $\bm{\mu} \in \mathbb{R}^n$ is the average of historical returns, and $\bm{\Sigma} \in \mathbb{R}^{n \times n}$ is the covariance matrix of historical returns. $\textit{turbulence}_t$ can be used to control buying/selling actions. If $\textit{turbulence}_t$ is higher than a preset threshold, the agent halts and will resume when $\textit{turbulence}_t$ becomes lower than the threshold.

%\subsubsection{Automated Backtesting}
%Backtesting plays a key role in evaluating the performance of a trading strategy. Automated backtesting tool is preferred because it reduces the human error. In the FinRL library, we use the Quantopian pyfolio package \cite{pyfolio} to backtest our trading strategies. It is easy to use and consists of various individual plots that provide a comprehensive image of the performance of a trading strategy.

\vspace{-0.1in}
\subsection{Baseline Strategies and Trading Metrics}
\label{sec:baseline_trading_strategies}

%\yanglet{To Bruce: Baselines for financial tasks should be well-chosen, what are the criteria? Why those? try to answer such questions.}

%Baseline trading strategies should be well-chosen and follow industrial standards. The strategies will be universal to measure, typical to compare with, and easy to implement.

Baseline trading strategies are provided to compare with DRL strategies. Investors usually have two mutually conflicting objectives: the highest possible profits and the lowest possible risks \cite{sharpe1970portfolio}. We include three conventional strategies as baselines.

\textbf{Passive trading strategy} \cite{Malkiel2003PassiveIS} is an easy and popular strategy that has the minimal trading activities. Investors simply buy and hold index ETFs \cite{Tarassov2016ExchangeTF} to replicate a broad market index or indices such as Dow Jones Industrial Average (DJIA) index and Standard \& Poor's 500 (S\&P 500) index.

%Popular index ETFs are: SPDR S\&P 500 ETF Trust (SPY) and Vanguard 500 Index Fund (VOO) for S\&P 500 Index, SPDR Dow Jones Industrial Average ETF Trust (DIA) for Dow Jones Industrial Average, Invesco QQQ Trust Series 1 (QQQ) for Nasdaq 100 index, etc.

\textbf{Mean-variance and min-variance strategy} \cite{meanvariance2012} both aim to achieve an optimal balance between the risks and profits. It selects a diversified portfolio with risky assets, and the risk is diversified when traded together.

\textbf{Equally weighted strategy} is a type of portfolio allocation method. It gives the same importance to each asset in a portfolio.

FinRL includes common metrics to evaluate trading performance:

%\noindent 
\noindent \textbf{Final portfolio value}: the amount of money at the end of the trading period.

\noindent \textbf{Cumulative return}: subtracting the initial value from the final portfolio value, then dividing by the initial value.

\noindent \textbf{Annualized return and standard deviation}: geometric average return in a yearly sense, and the corresponding deviation.

%\noindent \textbf{Annualized standard deviation}: annualized standard deviation of the portfolio return. % $R_{p}$.

\noindent \textbf{Maximum drawdown ratio}: the maximum observed loss from a historical peak to a trough of a portfolio, before a new peak is achieved. Maximum drawdown is an indicator of downside risk over a time period.  % during the trading period.
% FinRL pushes back at any historical point in the selected cycle, then find the nadir of total assets.

\noindent \textbf{Sharpe ratio} in (\ref{reward3}) is the average return earned in excess of the risk-free rate per unit of volatility.

\subsection{Hands-on Tutorials}
%逻辑上，是展示 对应的tutorial，描述实现任务的便利性 和展示回测结果是合理和靠谱的，开发上用户的便利
%应用层-指定什么东西，input到底是什么，拿出output是什么，系统自动配置，一键生成图。

We provide tutorials to help users walk through the strategy design pipeline, i.e., get familiar with the stat-action-reward specifications in Table \ref{event:eventTypes} and the agent-environment interactions in Fig. \ref{state_transition}.

\noindent \textbf{Tutorial 1: Stock trading}

%focus在education这个点，易上手的教学价值。
%强调的是教学价值和举的例子代表性
%输入很简单，输出很好分析
%trading on a time series  
%trading on a order book
%Analyzing trading performance

%We provide a tutorial for stock trading on time series data. 

First, users specify the \textit{state} at the application layer, i.e., the number of stocks, technical indicators, the initial capital, etc.  Second, users provide start/end dates for training/testing periods, set the time granularity. FinRL instantiates an environment for the task, while the operations are transparent to users. FinRL uses standard APIs to download data and obtains a Pandas DataFrame containing the open-high-low-close prices and volume (OHLCV) data. FinRL preprocesses the OHLCV data by filling missing data and calculates technical indicators that are passed into the state.  Third, users select a DRL library and a DRL algorithm. FinRL has default hyperparameters for daily stock trading task. During the testing period, users can tune these parameters to improve the trading performance. %The output is a set of actions including the positions to trade on for each step. 
Finally, FinRL feeds time series data of the portfolio value into a backtesting module to  plot charts. Please see examples in Section \ref{sec:stock_trading} and Section \ref{sec:portfolio_allocation}.

%\noindent \textbf{Tutorial 2: Bitcoin (BTC) Trading}

%Similar to Tutorial 1, we show how users can perform bitcoin (BTC) trading using FinRL.

%FinRL provides connection APIs to the CCXT platform \cite{CCXT}, which connects and trades with cryptocurrency exchanges and worldwide payment processing services. CCXT provides quick access to market data for analysis, indicator development, algorithmic trading, strategy backtesting, etc. 

\begin{figure*}
\centering
\includegraphics[width=0.95\textwidth]{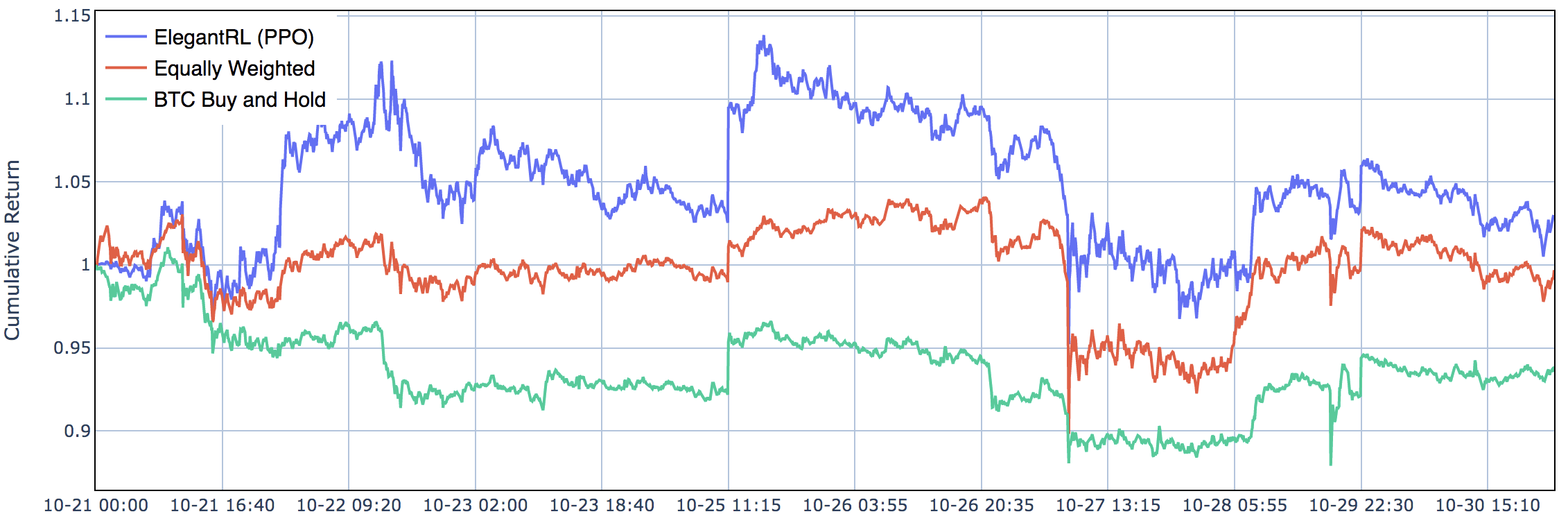}\vspace{-0.1in}
\caption{Cumulative returns (5-minute) of top 10 market cap cryptocurrencies trading using FinRL.}\vspace{-0.1in}
\label{portfolio_2}
\vspace{-0.1in}
\end{figure*}

%数据上用API和自动data preprocessing，便捷性
%env上可以直接用我们提供的成熟的gym的写法
%支持多个算法库，users可以更关注到底哪个算法更适合股票交易哪个算法更适合BTC交易，而不是重新复现这个算法。
%This tutorial designs an automated trading solution for stock trading. We present an PPO agent that trades stocks from the NASDAQ-100 constituents, such as AAPL, MSFT, AMZN, GOOGL, FB, which achieves the optimal policy in terms of expected revenues. The performance of stock trading is illustrated in Fig. 4 and Table 1.

%This tutorial utilizes three modules in our library: (i) command-line parsing files are used to combine the environment with the algorithm, (ii) it invokes a DRL algorithm module and the NASDAQ-100 dataset based environment to trade single stock, and (iii) it uses library functions for direct outputs. This use case demonstrates the convenience and utility of FinRL library functions.

\noindent \textbf{Tutorial 2: Analyzing Trading Performance}

Before deploying a trading strategy, users need to fully evaluate its trading performance via backtesting. The trading performance can be easily evaluated using the automatic backtesting module in Section \ref{sec:backtesting_module}. The commonly used trading metrics and baseline strategies are given in Section \ref{sec:baseline_trading_strategies}.

Cumulative return and Sharpe ratio are widely used metrics to evaluate overall performance of trading strategies. To gain more details about the strategy, the distribution of returns over the testing period and annualized return are provided to examine if the return is stable and consistent. Annualized volatility and maximum drawdown measure the robustness. 
%We can also use return distribution. If most of the annual returns are steady and consistent in terms of direction, and the returns have relatively high Sharpe ratio, then the returns on the portfolio over the exposed risk can be considered as good.
\vspace{-0.1in}
\subsection{Use Case I: Stock Trading}
\label{sec:stock_trading}

We use FinRL to reproduce both \cite{xiong2018practical} and \cite{yang2020} for stock trading. The ensemble strategy \cite{yang2020} combines three DRL algorithms (PPO \cite{schulman2017proximal}, A2C \cite{mnih2016asynchronous} and DDPG \cite{lillicrap2015continuous}) to improve the robustness. 

The implementation is easy with FinRL. We choose three algorithms (PPO, A2C, DDPG) in the agent layer, and an environment with start and end dates in the environment layer. 
%The application layer provides interface for users to specify trading settings and hyperparameters. 
The implementations of DRL algorithms and data preprocessing are transparent to users, alleviating the programming and debugging workloads. Thus, FinRL greatly facilitates the strategy design, allowing users to focus on improving the trading performance. 

Fig. \ref{stock_trading_1} and Table \ref{tab:Performance evaluation} show the backtesting performance on Dow 30 constituent stocks, accessed at 2020/07/01. The training period is from 2009/01/01 to 2020/06/30 on a daily basis, and the testing period is from 2020/07/01 to 2021/06/30. The performance in terms of multiple metrics is consistent with the results reported in \cite{yang2020} and \cite{xiong2018practical}, and here we show results in a recent trading period. We can see from the DJIA index that the trading period is a bullish market with an annual return of $32.84\%$. The ensemble strategy achieves a Sharpe ratio of $2.81$ and an annual return of $52.61\%$. It beats A2C with a Sharpe ratio of $2.24$, PPO with a Sharpe ratio of $2.23$, DDPG with a Sharpe ratio of $2.05$, DJIA with a Sharpe ratio of $2.02$, and min-variance portfolio allocation with a Sharpe ratio of $1.98$, respectively. Therefore, the backtesting performance demonstrates that FinRL successfully reproduces the ensemble strategy \cite{yang2020}.

\vspace{-0.1in}
\subsection{Use Case II: Portfolio Allocation}
\label{sec:portfolio_allocation}

%We show a reinforcement learning approach for portfolio allocation \cite{Jiang2017CryptocurrencyPM}. It learns a trading strategy to allocate money (assigned weights) on a set of stocks and reallocate periodically, say weekly. We use FinRL to provide backtesting results on the 30 Dow Jones stocks, as shown in Fig. X and Table X. 

%high level, plug-and-play,怎么用它，细节复现了，基于标准库复现的可读性很好。
%多app实现plug，突出架构和特性. The top layer includes applications in automated trading, where we demonstrate three use cases, namelystock trading, portfolio allocation and liquidation analysis.

We reproduce a portfolio allocation strategy \cite{Jiang2017ADR} that uses a DRL agent to allocate capital to a set of stocks and reallocate periodically. %, say weekly.  

FinRL improves the reproducibility by allowing users to easily compare the results of different settings, such as the pool of stocks to trade, the initial capital, and the model hyperparameters. It utilizes the agent layer to specify the state-of-the-art DRL libraries. Users do not need to redevelop the neural networks and instead they can just plug-and-play with any DRL algorithm. 
%FinRL has customized APIs to pass in data for different tasks, and users only need to pick the environment with start and end dates and do not need to work with API connection between programs and databases/platforms. 

Fig. \ref{portfolio_1} and Table \ref{tab:Performance evaluation} depict the backtesting performance on Dow 30 constituent stocks. The training and testing period is the same with Case I. It shows that each DRL agent, namely A2C \cite{mnih2016asynchronous}, TD3 \cite{fujimoto2018addressing}, PPO \cite{schulman2017proximal}, and DDPG \cite{lillicrap2015continuous}, outperforms the DJIA index and the min-variance strategy. A2C has the best performance with a Sharpe ratio of $2.36$ and an annual return of $42.57\%$; TD3 is the second best agent with a Sharpe ratio of $2.28$ and an annual return of $39.38\%$; PPO with a Sharpe ratio of $2.11$ and an annual return of $36.17\%$ and DDPG with a Sharpe ratio of $2.21$ and an annual return of $36.01\%$. Therefore, using FinRL, users can easily compare the agents' performance with each other and with the baselines.

\vspace{-0.1in}
\subsection{Use Case III:  Cryptocurrencies Trading}
\label{sec:portfolio_allocation}

We use FinRL to reproduce \cite{Jiang2017CryptocurrencyPM} for top 10 market cap cryptocurrencies \footnote{The top 10 market cap cryptocurrencies as of Oct 2021 are: Bitcoin (BTC), Ethereum (ETH), Cardano (ADA), Binance Coin (BNB), Ripple (XRP), Solana (SOL), Polkadot (DOT), Dogecoin (DOGE), Avalanche (AVAX), Uniswap (UNI).}. FinRL provides a full-stack development pipeline, allowing users to have an end-to-end walk-through of how to download market data using APIs, perform data preprocessing, pick and fine-tune DRL algorithms, and get automated backtesting performance.

Fig. \ref{portfolio_2} describes the backtesting performance on the ten cryptocurrencies with transaction cost. The training period is from 2021/10/01 to 2021/10/20 on a 5-minute basis, and the testing period is from 2021/10/21 to 2021/10/30. The portfolio with the PPO algorithm from the ElegantRL library has the highest cumulative return of $103\%$; Equally weighted portfolio strategy has the second highest cumulative return of $99\%$; BTC buy and hold strategy with a cumulative return of $93\%$. Therefore, the backtesting performance shows that FinRL successfully reproduce \cite{Jiang2017CryptocurrencyPM} with completeness and simplicity.

\vspace{-0.2in}
\section{Ecosystem of FinRL and conclusions}

In this paper, we have developed an open-source  framework, FinRL, to help quantitative traders overcome the steep learning curve. Customization is accessible on all layers, from market environments, trading agents up towards trading tasks. FinRL follows a training-testing-trading pipeline to reduce the simulation-to-reality gap. Within FinRL, historical market data and live trading platforms are reconfigured into standardized environments in OpenAI gym-style; state-of-the-art DRL algorithms are implemented for users to train trading agents in a pipeline; and an automated backtesting module is provided to evaluate trading performance. Moreover, benchmark schemes on typical trading tasks are provided as practitioners' stepping stones.

\noindent \textbf{Ecosystem of FinRL Framework}. We believe that the rise of the open-source community fostered the development of AI in Finance for both academia and industry side. As the need of utilizing open-source AI for finance ecosystem is imminent in the finance community, FinRL provides a ecosystem that features Deep Reinforcement Learning in finance comprehensively to fulfill such need for all-level users in our open-source community. 

FinRL offers an overall framework to utilize DRL agents for various markets, SOTA DRL algorithms, finance tasks (portfolio allocation, cryptocurrency trading, high-frequency trading), live trading support, etc. For entry-level users, FinRL aims to provide a demonstrative and educational atmosphere with hands-on documents to help beginners get familiar with DRL in Finance applications. For intermediate-level users, such as full-stack developers and professionals, FinRL provides ElegantRL \cite{erl}, a lightweight and scalable DRL library for FinRL with finance-oriented optimizations. For advanced-level users, such as investment banks and hedge funds. FinRL delivers FinRL-Podracer \cite{finrl_podracer_2021, gpu_podracer_nips_2021}, a cloud-native solution for FinRL with high performance and high scalability training.

FinRL also develops other useful tools to support the ecosystem. FinRL-Meta \cite{liu2021neofinrl} adds financial data engineering for FinRL with unified data processor and hundreds of market environments. Explainable DRL for portfolio management \cite{xfinrl_2021} and DRL ensemble strategy for stock trading \cite{xiong2018practical,yang2020} are also implemented. 

\noindent \textbf{Future work}.  Future research directions would be investiaging DRL's potential on limit order book \cite{vyetrenko2019get}, hedging \cite{buehler2019deep}, market making \cite{RL_market_making_2019}, liquidation \cite{bao2019multi}, and trade execution \cite{Lin2020ADR}.

\vspace{-0.10in}

\bibliographystyle{ACM-Reference-Format}
\bibliography{sections/ref.bib}

\end{document}